\documentclass{PoS}

\usepackage{units}

\title{Event-Shape Engineering and Muon--Hadron Correlations with ALICE}

\ShortTitle{Event-Shape Engineering and Muon--Hadron Correlations with ALICE}

\author{\speaker{Jan Fiete Grosse-Oetringhaus} for the ALICE collaboration\\
        CERN\\
        E-mail: \email{jgrosseo@cern.ch}}

\abstract{Angular correlations of two and more particles are a sensitive probe of
the initial state and the transport properties of the system produced in heavy-ion collisions.
Two recent results of the ALICE collaboration are presented. 
Event-shape engineering, a novel method, is applied to Pb--Pb collisions which splits events within the same centrality interval into classes with different average flow. The results indicate an interplay between radial and elliptic flow likely related to the initial-state eccentricity. \\
In pp and p--Pb collisions, recent results revealed intriguing long-range
correlation structures reminiscent of features observed in heavy-ion
collisions. The use of forward detectors allowed to show that long-range correlation structures persist also at large rapidities in p--Pb collisions.
}

\FullConference{The European Physical Society Conference on High Energy Physics\\
		22--29 July 2015\\
		Vienna, Austria}

\newcommand{\gevc}         {GeV/\ensuremath{c}}
\newcommand{\mevc}         {MeV/\ensuremath{c}}

\newcommand{\pt}           {\ensuremath{p_{\mathrm{T}}}{ }}

\newcommand{\snn}          {\ensuremath{\sqrt{s_{\mathrm{NN}}}}}

\newcommand{\Dphi}         {\ensuremath{\Delta\varphi}}
\newcommand{\Deta}         {\ensuremath{\Delta\eta}}

\newcommand{\com}[1]       {}

\begin{document}


These proceedings present two recent results from the ALICE collaboration giving further insight into collective effects in Pb--Pb and p--Pb collisions. 
For the first result a novel technique, event-shape engineering, is applied to learn about the correlation between the initial state of the collision and the hydrodynamical expansion. 
The second result addresses the question if the intriguing ridge structures observed in high multiplicity p--Pb collisions range out to forward rapidities.

\section{Event-Shape Engineering}

\begin{figure}[b]
  \centering
  \includegraphics[width=0.95\textwidth]{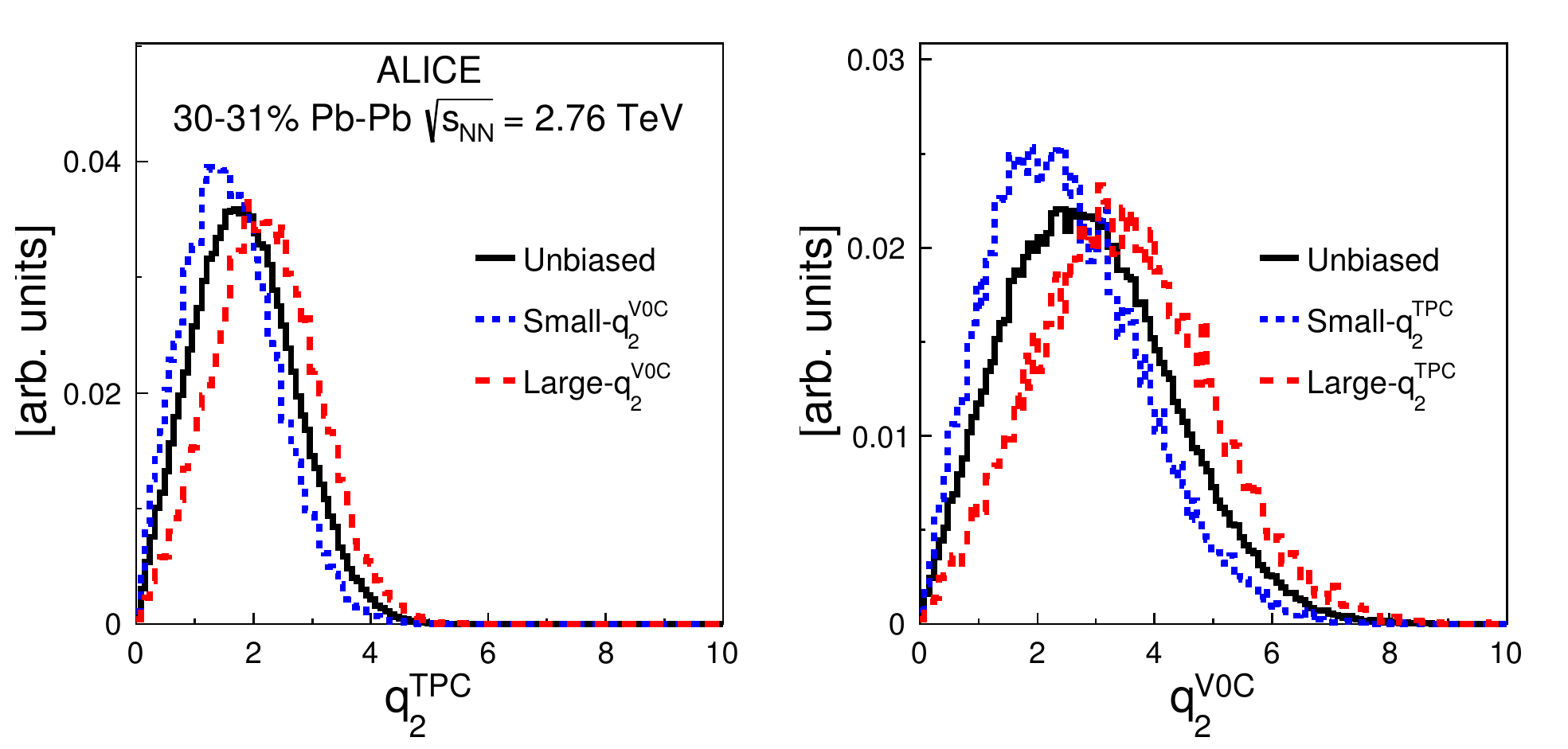}
  \caption{\label{fig:qvectorselection} Distribution of $q_2^{\rm TPC}$ when selecting on $q_2^{\rm V0C}$ (left panel) and $q_2^{\rm V0C}$ when selecting on $q_2^{\rm TPC}$ (right panel). Figures from \cite{Adam:2015eta}.}
\end{figure}

\begin{figure}[t]
  \centering
  \includegraphics[width=\textwidth]{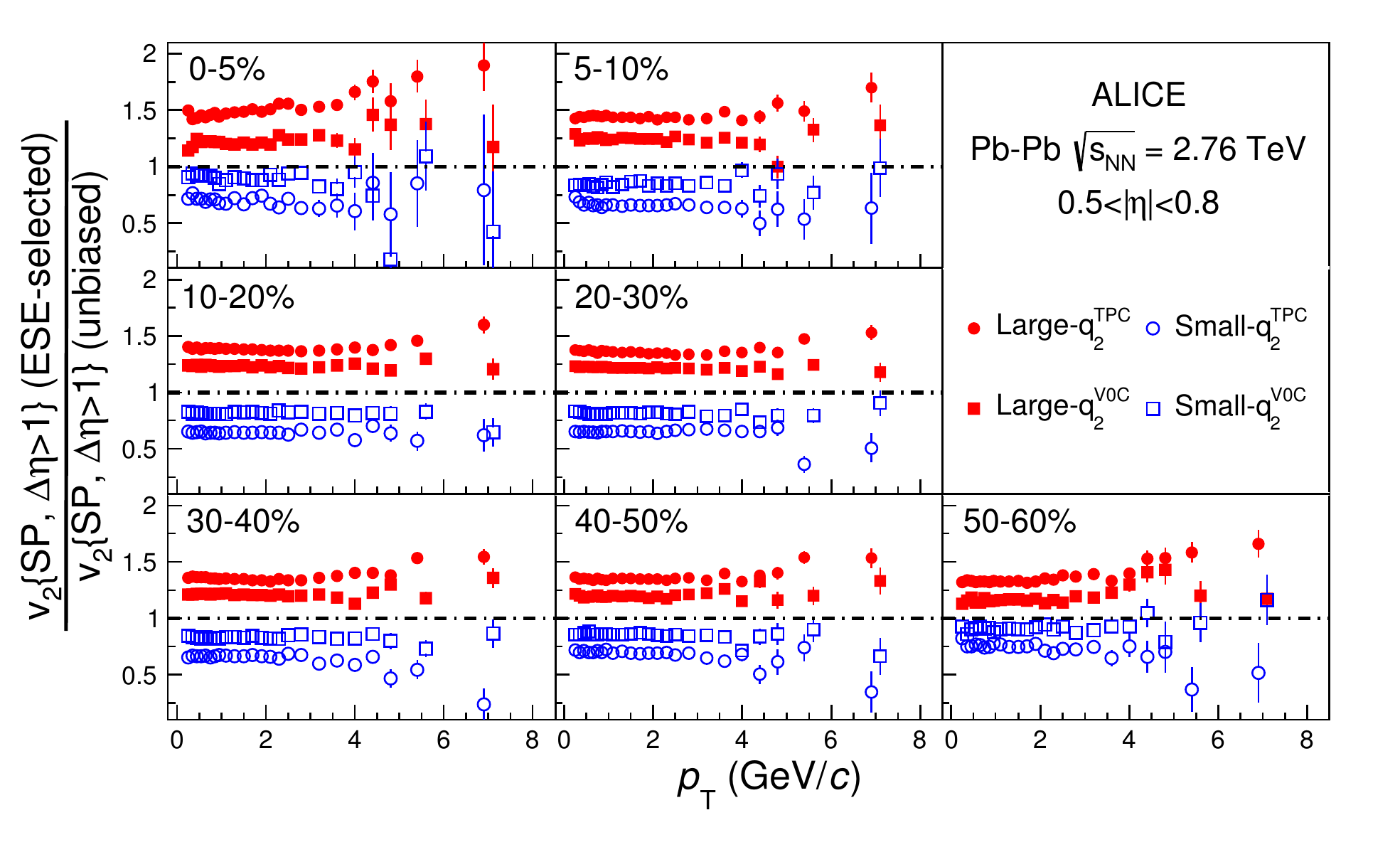}
  \caption{\label{fig:v2ratio} Ratio of $v_2{\rm\{SP\}}$ with $q_2$ selection over inclusive events for centrality classes ranging from 0 to 60\%. Figure from \cite{Adam:2015eta}.}
\end{figure}

The method of event-shape engineering \cite{Schukraft:2012ah} exploits the large event-by-event variation of the $v_n$ coefficients \cite{Abelev:2012di} which are correlated to the initial-state eccentricities \cite{Heinz:2013th}. The idea is to split the events based on their initial geometry and study the effect on final-state observables in the event. In the analysis \cite{Adam:2015eta} presented in these proceedings events are split by the event-by-event second-order reduced flow vector $q_2$:
\begin{equation}
  q_2 = \frac{\sqrt{Q_{2,x}^2 + Q_{2,y}^2}}{\sqrt{M}} \mbox{  with  } Q_{2,x} = \sum_i \cos 2 \varphi_i \mbox{  and  } Q_{2,y} = \sum_i \sin 2 \varphi_i
\end{equation}
where $M$ is the event multiplicity and $i$ runs over all charged particles.
$q_2$ is determined in two different phase-space regions: $|\eta| < 0.4$ called $q_2^{\rm TPC}$ and $-3.7 < \eta < -1.7$ called $q_2^{\rm V0C}$. Selecting $q_2$ in one region and studying $q_2$ in the other region (Fig.~\ref{fig:qvectorselection}) shows that $q_2$ is correlated over several units of pseudorapidity. 
Tracks from $0.5 < |\eta| < 0.8$ are used to study the elliptic flow $v_2{\rm\{SP\}}$ using the scalar-product method and $\pt$ spectra.
Figure~\ref{fig:v2ratio} presents the ratio of the $\pt$-differential $v_2{\rm\{SP\}}$ after $q_2$ selection over the inclusive sample. It can be seen that $v_2{\rm\{SP\}}$ is larger for large $q_2$ and smaller for small $q_2$ as expected. In addition, the ratio is independent of $\pt$ suggesting that the $q_2$ selection actually acts on a global event property.

\begin{figure}[t]
  \centering
  \includegraphics[width=0.49\textwidth]{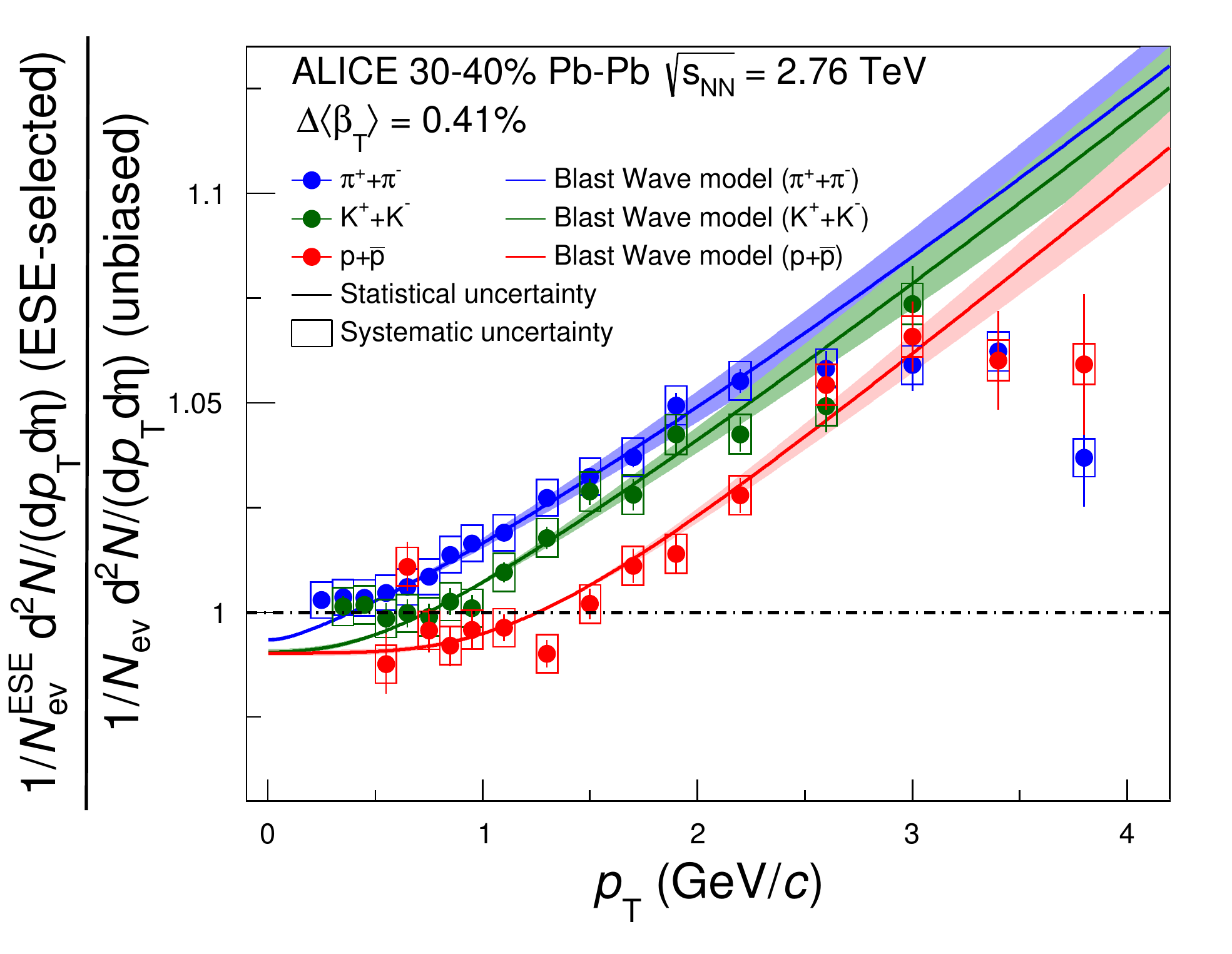}
  \hfill
  \includegraphics[width=0.49\textwidth]{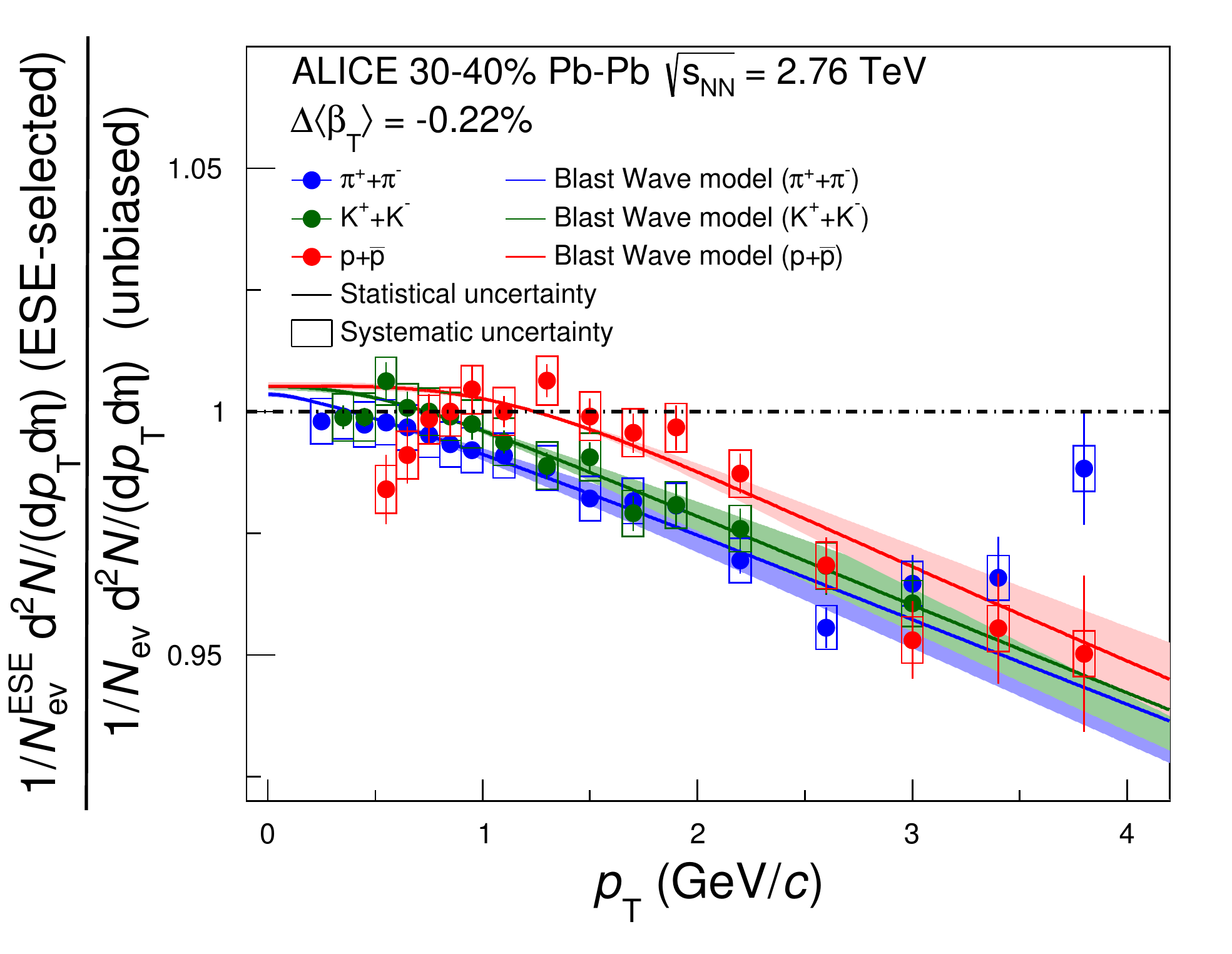}
  \caption{\label{fig:blastwave} Blast-wave fits to the ratio of $\pt$ spectra over inclusive events for large (left panel) and small $q_2$ (right panel). The points show the data while the lines show the fit. Figures from \cite{Adam:2015eta}.}
\end{figure}

The next step is to study if other properties of the event change with the $q_2$ selection. Figure~\ref{fig:blastwave} presents the change to $\pt$-spectra of $\pi$, K, and p. One observes harder (softer) spectra for large (small) values of $q_2$. The change in the spectra depends on the type of the particle. To quantify this effect a blast-wave fit~\cite{Schnedermann:1993ws} is performed. In the fit the temperature is kept fixed to the value of the inclusive spectra~\cite{Abelev:2013vea}, while the radial expansion velocity $\beta_T$ is allowed to change. The fit describes the particle-species dependence well and a larger (smaller) $\beta_T$ is found for large (small) $q_2$. This suggests a positive correlation of $q_2$ (which can be related to the initial-state eccentricity) with the radial expansion (which is related to the radial pressure gradient). These results will further constrain the initial-state and hydrodynamical modeling. More details can be found in \cite{Adam:2015eta}.

\section{Forward Muon--Hadron Correlations}

\begin{figure}[t]
  \centering
  \hspace{0.5cm}
  \includegraphics[width=0.4\textwidth]{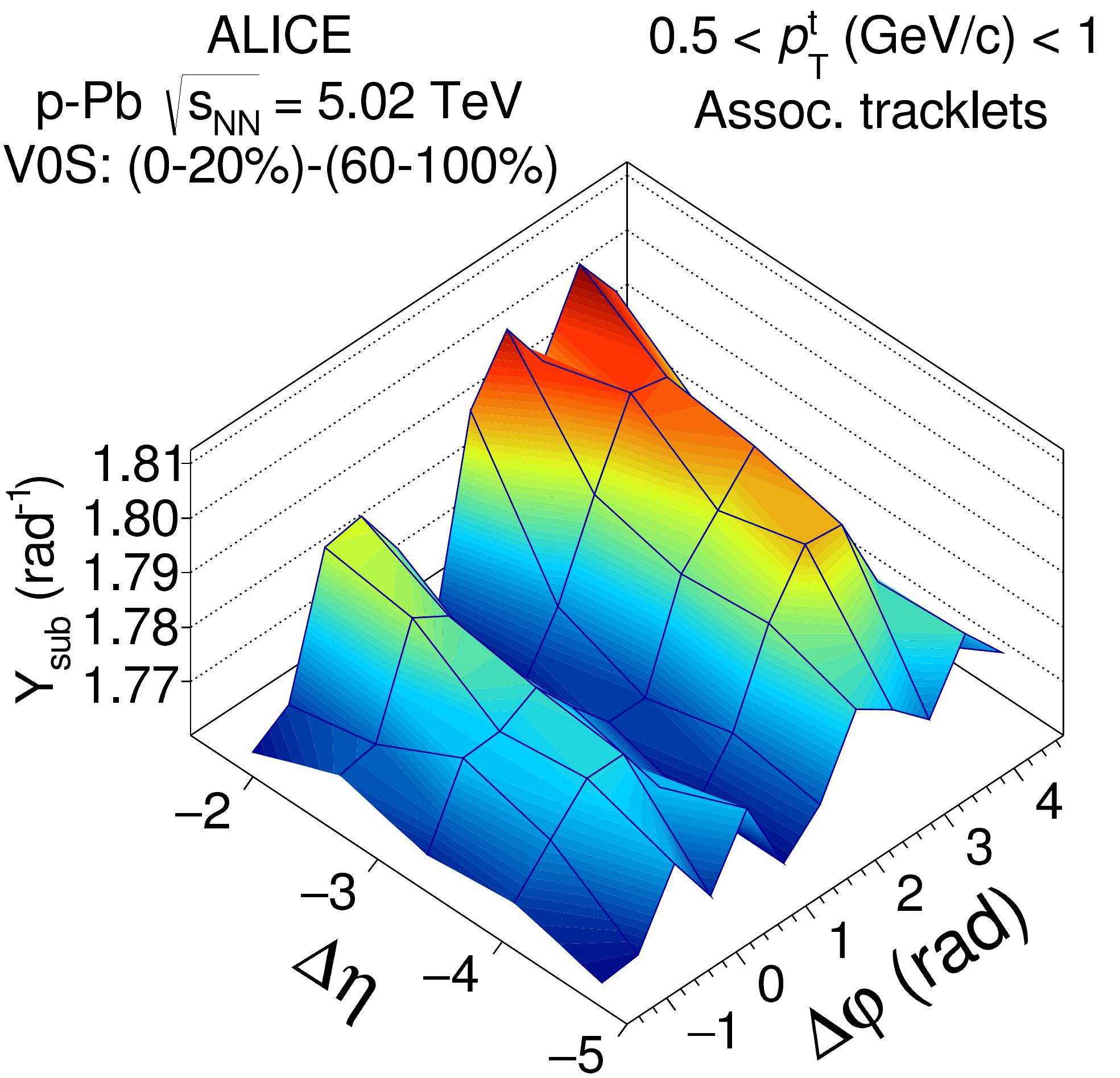}
  \hfill
  \includegraphics[width=0.4\textwidth]{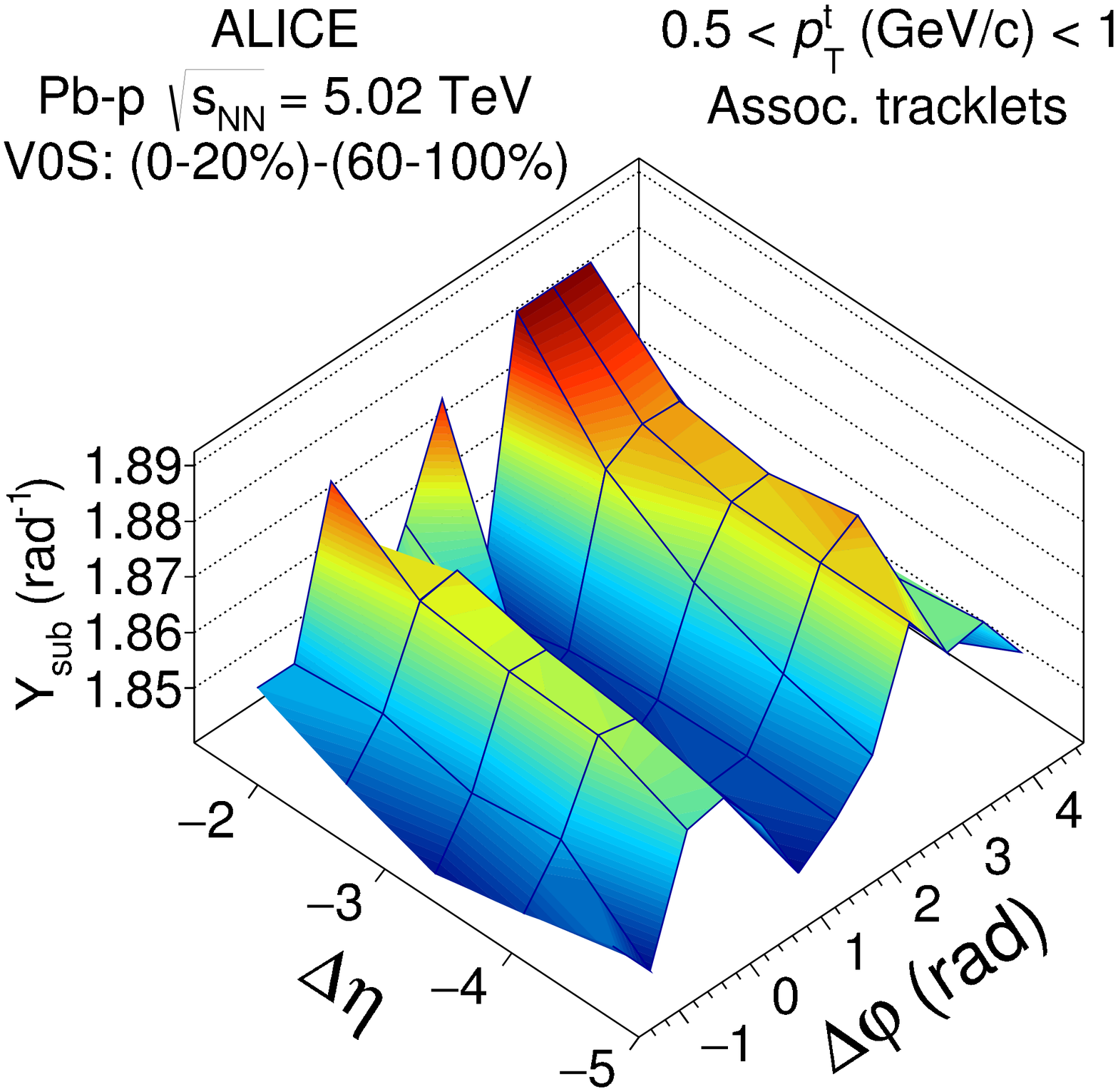}
  \hspace{0.5cm}
  \caption{\label{fig:ridge} Two-particle correlations in p-going (left panel) and Pb-going (right panel) direction for high-multiplicity collisions after the subtraction of low-multiplicity collisions. Figures from \cite{Adam:2015bka}.}
\end{figure}


The intriguing discovery of the near-side ridge in high-multiplicity p--Pb collisions \cite{CMS:2012qk} and a second ridge on the away side \cite{alice_pa_ridge,atlasridge} resulted in significant experimental and theoretical work to study collective effects in small systems. Nowadays, the collective nature of the observed features in p--Pb collisions is rather well established based on a number of observations: the correlation structures show a typical particle-species dependence \cite{ABELEV:2013wsa} and involve at least 8 particles \cite{Khachatryan:2015waa}. 

\begin{figure}[t]
  \centering
  \hspace{0.5cm}
  \includegraphics[width=0.44\textwidth]{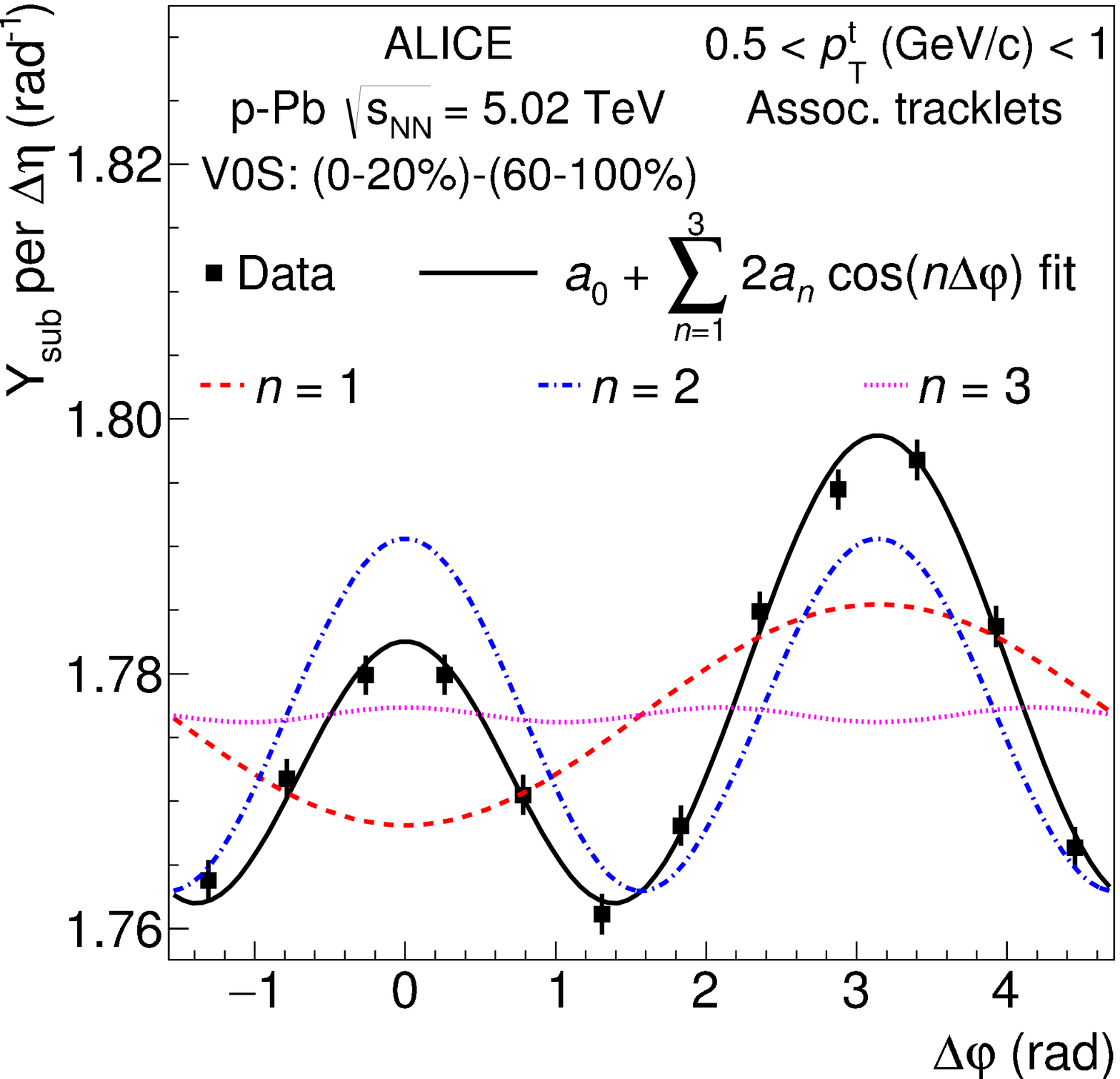}
  \hfill
  \includegraphics[width=0.44\textwidth]{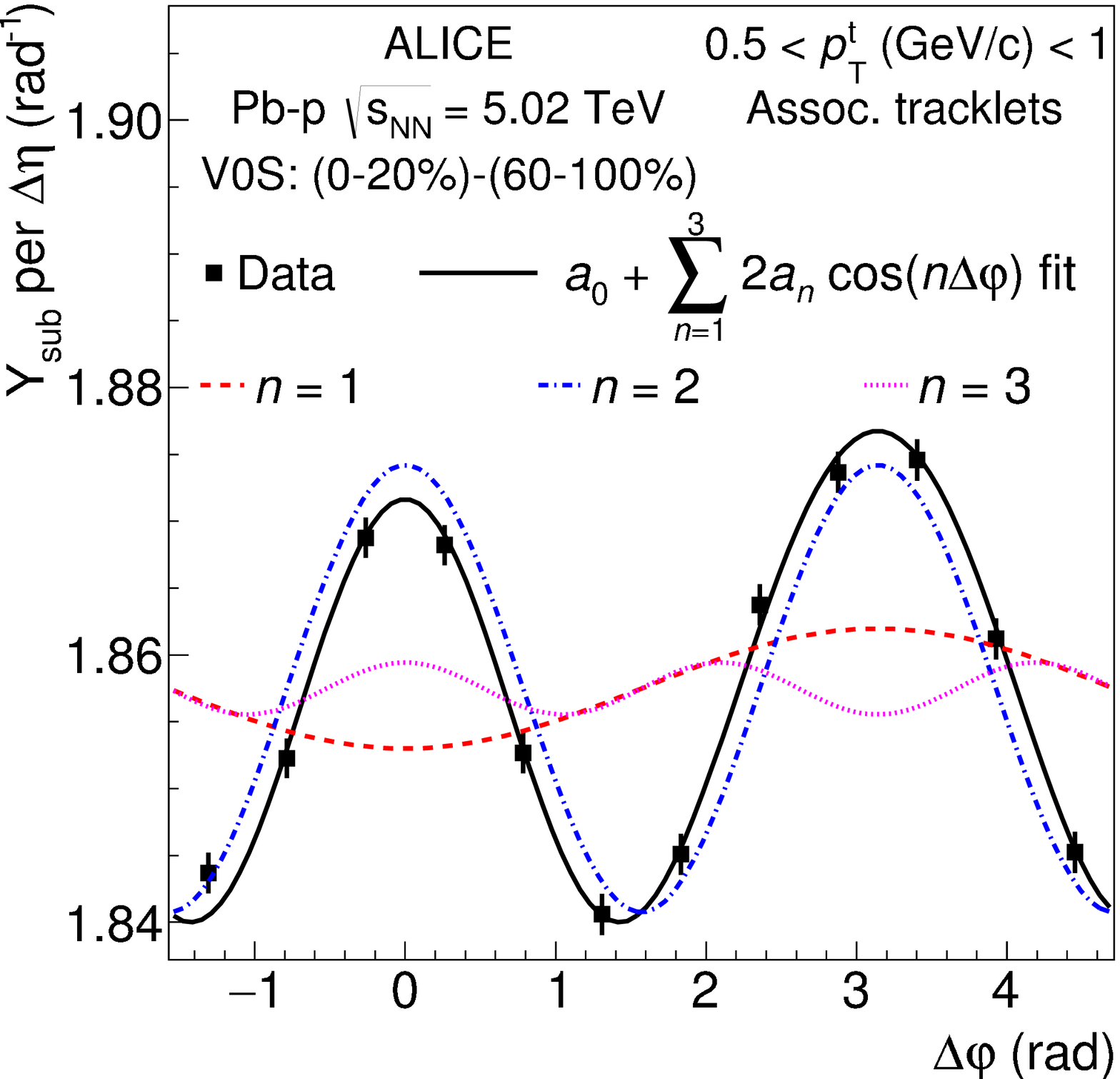}
  \hspace{0.5cm}
  \caption{\label{fig:ridge_projections} Projections of two-particle correlations to $\Dphi$ in p-going (left panel) and Pb-going (right panel) direction for high-multiplicity collisions after the subtraction of low-multiplicity collisions. The lines indicate the first three Fourier components of the distribution. Figures from \cite{Adam:2015bka}.}
\end{figure}

The results reported in these proceedings study this structure at forward rapidities \cite{Adam:2015bka}. Muons measured with the forward muon spectrometer ($-4 < \eta < -2.5$) are correlated with tracklets measured at mid-rapidity ($|\eta| < 1$). Tracklets are reconstructed using the two innermost layers of the ALICE Inner Tracking System. Particles with a $\pt$ larger than about \unit[50]{\mevc} can be reconstructed in this way.
The measured muons contain contributions from pion and kaon decays which dominate below a $\pt$ of about \unit[1.5]{\gevc}. Above \unit[2]{\gevc} the muons stem mostly from heavy-flavor decays.
The LHC collided p--Pb collisions at $\snn = \unit[5.02]{TeV}$ in two different beam configurations. The proton was either going towards the muon spectrometer (called p-going direction in the following) or in the opposite direction, thus the lead ion was going in direction of the muon spectrometer (called Pb-going direction). The center of mass system is shifted with respect to the lab system by $\pm 0.465$ depending on beam direction. 
Events are split into multiplicity classes based on the combined multiplicity measured in $2.8 < \eta < 3.9$ and $-3.7 < \eta < -2.7$ denoted by \emph{V0S}. Events with the 20\% highest multiplicity are referred to with 0--20\% in the following while 60--100\% refers to events with the 40\% lowest multiplicity.
Two-particle correlations are measured as a function of azimuthal difference $\Dphi$ and pseudorapidity difference $\Deta$. In order to reduce the jet contribution, the two-particle correlation in the 60--100\% event class is subtracted from the one in 0--20\% event class \cite{alice_pa_ridge}. 
Figure~\ref{fig:ridge} presents two-particle correlations for p-going and Pb-going direction. A double-ridge structure is observed in both directions up to $\Deta \approx 5$, thus out to $\eta \approx 4$. Projections of these correlations to $\Dphi$ are shown in Fig.~\ref{fig:ridge_projections} together with their Fourier decomposition:
\begin{equation}
  Y_{\rm sub} = a_0 + \sum_{n=1}^3 2 a_n \cos n \Dphi.
\end{equation}
The second coefficient dominates in both beam directions. From the $a_2$ coefficient the relative modulation $V_{2\Delta}^{\mu-h}$ in the 0--20\% event class can be calculated. Assuming that this modulation is caused by a modulation of the single-particle distribution with respect to a common symmetry plane, the corresponding coefficients \cite{Aamodt:2011by} can be calculated with $v_2^\mu{\rm\{2PC,sub\}} = V_{2\Delta}^{\mu-h} / \sqrt{V_{2\Delta}^{h-h}}$ where $V_{2\Delta}^{h-h}$ is from two-particle correlations at mid-rapidity.

\begin{figure}[t]
  \centering
  \includegraphics[width=0.6\textwidth]{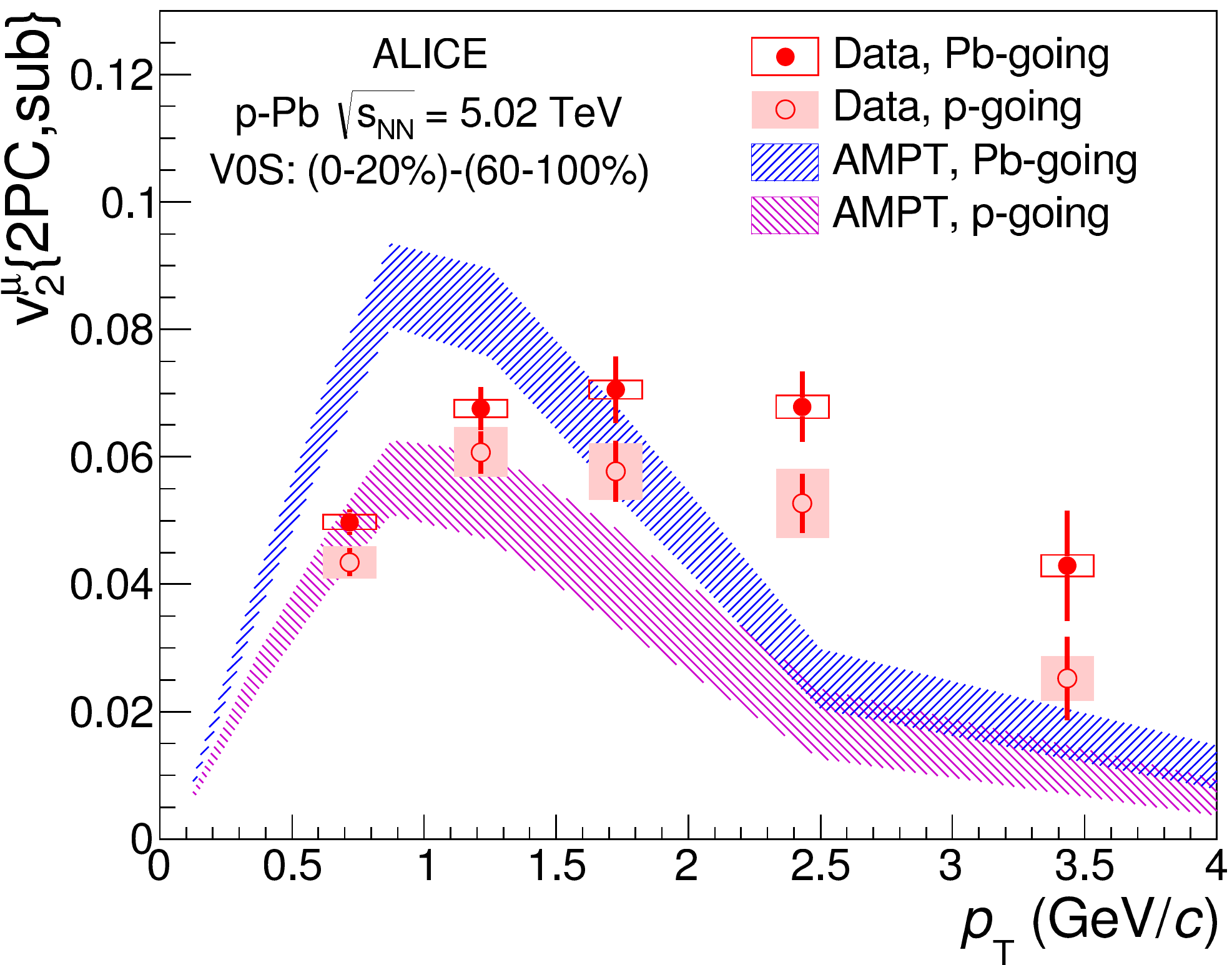}
  \caption{\label{fig:ridge_v2} $v_2^\mu{\rm\{2PC,sub\}}$ coefficient extracted from muon--hadron correlations after low-multiplicity subtraction (for details see text). Open symbols show p-going direction, while filled symbols are for Pb-going direction. The result is compared to AMPT simulations. Figure from \cite{Adam:2015bka}.}
\end{figure}

Figure~\ref{fig:ridge_v2} presents the extracted $v_2^\mu{\rm\{2PC,sub\}}$ as a function of $\pt$. The values are found to be slightly larger for Pb-going than for p-going direction possibly related to an event-plane decorrelation in $\eta$-direction \cite{Khachatryan:2015oea}. 
A comparison to the AMPT model \cite{Lin:2004en, Xu:2011fi} is shown which follows the trend for low $\pt$ but underestimates the data at higher $\pt$. This is interesting as at higher $\pt$ the muons are dominated by heavy-flavor decays while the $v_2$ of such muons in AMPT is zero. Thus this can be understood as a hint for a non-zero heavy-flavour $v_2$ in the data or a largely different particle composition.
More details about these results can be found in \cite{Adam:2015bka}.

\bibliographystyle{pos}
\bibliography{biblio}{}

\begin{thebibliography}{10}

\bibitem{Adam:2015eta}
J.~Adam {\em et~al.}, {\em {Event shape engineering for inclusive spectra and
  elliptic flow in Pb-Pb collisions at $\sqrt{s_{\rm NN}}=2.76$ TeV},} {\em
  arXiv:1507.06194}, 2015.

\bibitem{Schukraft:2012ah}
J.~Schukraft, A.~Timmins, and S.~A. Voloshin, {\em {Ultra-relativistic nuclear
  collisions: event shape engineering},} {\em Phys. Lett.}, vol.~B719,
  pp.~394--398, 2013.

\bibitem{Abelev:2012di}
B.~Abelev {\em et~al.}, {\em {Anisotropic flow of charged hadrons, pions and
  (anti-)protons measured at high transverse momentum in Pb--Pb collisions at
  $\snn$=2.76 TeV},} {\em Phys.Lett.B}, vol.~719, pp.~18--28, 2013.

\bibitem{Heinz:2013th}
U.~Heinz and R.~Snellings, {\em {Collective flow and viscosity in relativistic
  heavy-ion collisions},} {\em Ann. Rev. Nucl. Part. Sci.}, vol.~63,
  pp.~123--151, 2013.

\bibitem{Schnedermann:1993ws}
E.~Schnedermann, J.~Sollfrank, and U.~W. Heinz, {\em {Thermal phenomenology of
  hadrons from 200-A/GeV S+S collisions},} {\em Phys. Rev.}, vol.~C48,
  pp.~2462--2475, 1993.

\bibitem{Abelev:2013vea}
B.~Abelev {\em et~al.}, {\em {Centrality dependence of $\pi$, K, p production
  in Pb-Pb collisions at $\sqrt{s_{NN}}$ = 2.76 TeV},} {\em Phys. Rev.},
  vol.~C88, p.~044910, 2013.

\bibitem{Adam:2015bka}
J.~Adam {\em et~al.}, {\em {Forward-central two-particle correlations in p-Pb
  collisions at $\sqrt{s_{\rm NN}}$ = 5.02 TeV},} {\em arXiv:1506.08032}, 2015.

\bibitem{CMS:2012qk}
S.~Chatrchyan {\em et~al.}, {\em {Observation of long-range near-side angular
  correlations in proton-lead collisions at the LHC},} {\em Phys. Lett.},
  vol.~B718, pp.~795 -- 814, 2013.

\bibitem{alice_pa_ridge}
B.~Abelev {\em et~al.}, {\em {Long-range angular correlations on the near and
  away side in p--Pb collisions at $\snn$=5.02 TeV},} {\em Phys.Lett.},
  vol.~B719, pp.~29--41, 2013.

\bibitem{atlasridge}
G.~Aad {\em et~al.}, {\em {Observation of associated near-side and away-side
  long-range correlations in $\snn$=5.02 TeV proton-lead collisions with the
  ATLAS detector},} {\em Phys.Rev.Lett.}, vol.~110, p.~182302, 2013.

\bibitem{ABELEV:2013wsa}
B.~B. Abelev {\em et~al.}, {\em {Long-range angular correlations of $\rm \pi$,
  K and p in p-Pb collisions at $\sqrt{s_{\rm NN}}$ = 5.02 TeV},} {\em Phys.
  Lett.}, vol.~B726, pp.~164--177, 2013.

\bibitem{Khachatryan:2015waa}
V.~Khachatryan {\em et~al.}, {\em {Evidence for Collective Multiparticle
  Correlations in p-Pb Collisions},} {\em Phys. Rev. Lett.}, vol.~115, no.~1,
  p.~012301, 2015.

\bibitem{Aamodt:2011by}
K.~Aamodt {\em et~al.}, {\em {Harmonic decomposition of two-particle angular
  correlations in Pb--Pb collisions at $\snn = 2.76$ TeV},} {\em Phys.Lett.},
  vol.~B708, pp.~249--264, 2012.

\bibitem{Khachatryan:2015oea}
V.~Khachatryan {\em et~al.}, {\em {Evidence for transverse momentum and
  pseudorapidity dependent event plane fluctuations in PbPb and pPb
  collisions},} {\em Phys. Rev.}, vol.~C92, no.~3, p.~034911, 2015.

\bibitem{Lin:2004en}
Z.-W. Lin, C.~M. Ko, B.-A. Li, B.~Zhang, and S.~Pal, {\em {A Multi-phase
  transport model for relativistic heavy ion collisions},} {\em Phys.Rev.},
  vol.~C72, p.~064901, 2005.

\bibitem{Xu:2011fi}
J.~Xu and C.~M. Ko, {\em {Pb-Pb collisions at $\sqrt{s_{NN}}=2.76$ TeV in a
  multiphase transport model},} {\em Phys.Rev.}, vol.~C83, p.~034904, 2011.

\end{thebibliography}

\end{document}